\documentclass{iopart}

\usepackage{graphicx}
\usepackage{url}
\usepackage{array}
\begin{document}

\title[Enhancement of absorption bistability by
trapping light planar metamaterial] {Enhancement of absorption
bistability by trapping light planar metamaterial}

\author{Vladimir R Tuz$^{1,2}$, Valery S Butylkin$^3$ and Sergey L Prosvirnin$^{1,2}$}

\address{$^1$ Institute of Radio Astronomy of National Academy of Sciences of Ukraine,
Kharkiv 61002, Ukraine}
\address{$^2$ School of Radio Physics, Karazin Kharkiv National University, 4,
Svobody Square, Kharkiv 61077, Ukraine}
\address{$^3$ Kotelnikov Institute of Radioengineering and
Electronics of Russian Academy of Sciences, Fryazino Branch, 1,
Vvedenskogo Square, Fryazino, Moscow Region, 141190, Russia}
\ead{tvr@rian.kharkov.ua, vasebut@ms.ire.rssi.ru and
prosvirn@rian.kharkov.ua}

\begin{abstract}
We propose to achieve a strong bistable response of a thin layer of
a saturable absorption medium by involving a planar metamaterial
specially designed to bear a high-Q trapped-mode resonance in
the infrared region.
\end{abstract}
\submitto{\JOA} \maketitle

\section{Introduction}

The effect of optical bistability (or multistability) is a basis of
numerous applications such as optical switching, differential
amplification, unidirectional transmission, power limiting, pulse
shaping, and optical digital data processing \cite{gibbs}. A common
way to obtain optically bistable devices is by using a resonant cavity
containing a Kerr-type or saturable absorption material whose
complex refraction index depends on the input light intensity. In
the first case, the \textit{dispersion} bistability appears as a
result of the nonlinear increasing or decreasing of the optical path
length as the field intensity inside the resonator rises. In the
second one, the dependence of power absorption of the material
inside the cavity on the input field intensity leads to the
saturable \textit{absorption} bistability. In the present paper we
mainly focus our attention on the latter phenomenon.

The mechanism of saturable absorption is related to the possibility
of a material absorbing more than one photon before relaxing to the
ground state. Also, two photons and multiphotons absorptions are
known under the sufficiently high light intensity. In addition,
population redistribution induced by intense laser fields leads to
interaction of stimulated emission and absorption in complex
molecular systems, and the generation of free carriers in solids
\cite{sutherland,christ}. These phenomena are manifested in the both
reduced (saturable) and increased (reverse saturable) optical
absorption.

The basic principle of obtaining optical bistable switching consists
in using a saturable absorber to detune a resonant cavity by
variation of the input light intensity \cite{gibbs}, \cite{szoke}. At
low intensity, the saturable absorber is characterized by high
attenuation which detunes the cavity and, thereby, causes the cavity
to reflect and absorb substantially all the incident wave power. At an
intensity above a threshold level of the system, the attenuation
decreases abruptly, and the cavity transmits substantially all the
incident wave energy. As light intensity is decreased, the cavity
continues to transmit light until a lower threshold level is
reached, at which point the medium again becomes a high absorber and
the cavity once again is detuned. The described process is
characterized by dependence between the input and output intensities
in the form of a hysteresis loop.

The threshold levels and the total change in attenuation depend upon
a kind of the material. The key parameters for a saturable
absorption material are its wavelength range (where it absorbs), its
dynamic response (how fast it recovers), and its saturation
intensity and fluence \cite{haiml}. Semiconductor materials,
combined with proper epitaxial growth and correct optical design of
the structure, can achieve a broad range of desired properties for
nearly ideal saturable absorber structures for all solid-state
lasers. Also the glasses doped by rare-earth elements or
semiconductor quantum dots (QD's) have also been proposed as
saturable absorbers to obtain optical switching in the IR range
\cite{malyarevich,zakery}.

Typically a saturable absorber is integrated inside a Fabry-Perot
structure or an optical ring cavity \cite{joshi_physrev,
joshi_physrev-2004}. In the further realizations the intracavity
saturation absorber is integrated in a more general mirror structure
that allows for both saturable absorption and negative dispersion
control, which is now generally referred to as semiconductor saturable
absorber mirrors (SESAM's) \cite{keller}. In a general sense the
design problem of SESAM's is reduced to the analysis of multilayered
interference filters for a desired nonlinear reflectivity response.
But in any case, some resonant cavity is used to provide the field
confinement and the nonlinear response enhancement.

A perspective way to reduce the cavity size and the input field
intensity required for optical bistable switching lies in using
plasmonic and metamaterial technologies \cite{sarychev-2007}. It is
known that the planar metamaterials can create an environment
equivalent to a resonant cavity which allows us to achieve the required
field confinement within a system to provide the nonlinear effect
enhancement. To date there are a set of publications
\cite{brien,pendry,gorkunov,lapine,soukoulis,poutrina,shadrivov}
where the properties of nonlinear metamaterials formed by
integrating nonlinear components or materials into the metamaterial
are studied both theoretically and experimentally. Typically such
structures are composed of metallic inclusions in the form of
\textit{symmetrical} split-ring resonators which are resonant
because of an internal capacitance and inductance within each
element. These elements are made nonlinear and tunable via the
insertion of diodes with a voltage-controlled characteristic in the
capacitive gaps of the metamaterial elements. To the best of our knowledge, in this
way only a dispersion bistable response in the metamaterials is
realized. A straightforward homogenization procedure is applied to
describe such nonlinear systems as a composite medium with
expression its material properties via some effective parameters. It
should be noted that the main drawbacks of such structures are the
low quality factor of resonances and considerable technological
difficulties of their manufacturing in the optical range.

Hence the metamaterials which are capable of supporting the
trapped-mode resonant excitation \cite{sprosvirnin,zouhdi,marinica}
can be a successful alternative. The trapped-mode resonances appear
in planar metamaterials designed on the basis of multi-element
periodic arrays which typically consist of identical subwavelength
metallic inclusions structured in the form of double-rings (DR's)
\cite{papasimakis} or \textit{asymmetrically} split rings or squares
\cite{fedotov,khardikov}. These particles are arranged periodically
and placed on a thin dielectric substrate. The trapped-mode
resonance is a result of the \textit{antiphase} current oscillations
in the particles of a periodic cell and they have high quality
factor due to the weak interaction with free space. The nonlinearity can
be included in such a system in the form of a substrate made of some
Kerr-type or saturable absorption nonlinear material.

This conception was recently confirmed experimentally in
\cite{tanaka,samson}, where the metamaterials which bear
trapped-mode resonances were proposed to realize an enhancement of
QD luminescence and electro-optic switching. The main advantage of
the studied structures is the fact that a significant field
enhancement is achieved in a thin planar system. Note that in these
structures an active medium was used as a substrate of planar
metamaterial. Also in our previous publications \cite{tuz,tuzv} the
features of a dispersion bistable response of metamaterials with
asymmetrical and symmetrical particles placed on a nonlinear
substrate were studied. It is shown that the effect of nonlinearity
appears as the formation of closed loops of bistable transmission
within the frequency of the trapped-mode resonance due to the strong
field localization in the system.

The goal of this paper is to show an enhancement of absorption
bistability in a planar metamaterial which bears the trapped-mode
resonances in sight to design the improved saturable absorbtion
mirror. An array of metallic double-ring elements is chosen to
construct the metamaterial due to the \textit{polarization
insensitivity} of this structure at the normal incidence of an
exciting wave.

\section{Problem statement and method of solution}

\begin{figure}[htb]
\centerline{\includegraphics[width=0.6\linewidth]{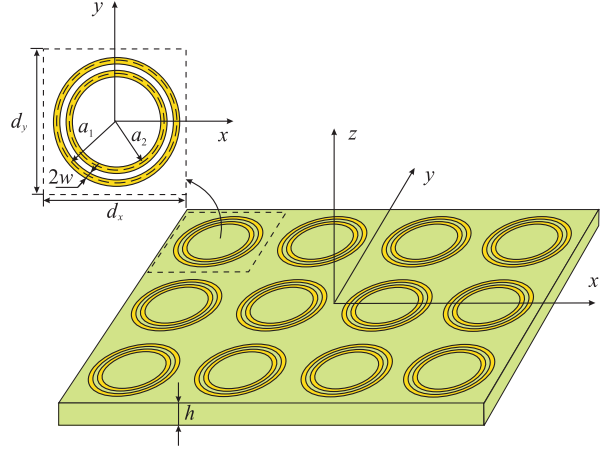}}
\caption{(Color online) Fragment of the planar metamaterial of DR's
and its elementary unit cell.} \label{fig:fig1}
\end{figure}

The square unit cell of the structure under study has a size
$d=d_x=d_y=800$~nm and consists of one metallic DR
(Fig.~\ref{fig:fig1}). The radii of the outer and inner rings are
fixed at $a_1=290$~nm and $a_2=230$~nm, respectively. The width of
the both rings is $2w=40$~nm. The array is placed on a dielectric
substrate with permittivity $\varepsilon$ and thickness $h=150$~nm.
Suppose that the normally incident plane monochromatic wave has a
frequency $\omega$ and a magnitude $A$.

When the intensity of input light is small, i.e. in the
\textit{linear regime}, in order to calculate the optical response
of  metamaterial, two different techniques based on the method of
moments (MoM) \cite{prosvirnin-1999-tpw} and the pseudo-spectral
time-domain (PSTD) method \cite{khardikov} were proposed earlier.
The MoM is a numerical-analytical method that involves solving an
integral equation which is related to the surface currents induced
on the metallic pattern by the incident electromagnetic wave, then
calculating the scattered fields produced by the currents as a
superposition of partial spatial waves. In the framework of this
method the metallic pattern is treated as a thin perfect conductor.
On the other hand, the PSTD is a direct numerical scheme which uses
a spatial spectral expansion of fields and the differentiation
theorem for Fourier transforms to calculate the spatial derivatives.
It is significant that the PSTD allows us to take into account the
strong dissipation and dispersion of the metal permittivity of the
elements. Nevertheless, in \cite{khardikov} it was revealed that the
results obtained with the both techniques are in good agreement
with each other down to the mid-IR region where the trapped-mode
resonances are still well observed and have high Q factor.

So we restrict our study within this region and use the MoM because
it allows us to calculate the magnitude of currents $J$ along the DR
element since it is important to estimate the inner field magnitude
inside the system. On the other hand, such a numerical-analytical
treatment enables us to clearly understand the physics of the
phenomena under investigation. Thus, in virtue of the MoM, the
magnitude of the current $J$, the reflection $R$, transmission $T$
and absorbtion $W=1-|R|^2-|T|^2$ coefficients can be determined
numerically as functions of frequency and permittivity of substrate
and presented in the symbolic form
\begin{equation}\label{eq:linear}
u=F_u(\omega,\varepsilon),
\end{equation}
where $u=J,R,T$, or $W$.

Next we should note that, at the trapped-mode resonance, due to
specific current distribution on the metallic pattern, the
electromagnetic energy is symmetrically confined to a very small
region between two half-arcs of the rings, where the energy density
reaches substantially high values. The latter feature allows us to
construct approximate model based on the transmission line theory to
estimate the inner field intensity caused by rings \cite{tuzv}.
According to this theory, in each unit cell, the DR is symmetrically
divided into two parts where the half-arc of the inner and outer
conductive rings are considered as two wires with a distance
$b=a_1-a_2-2w$ between them. Along these wires the currents with
equal magnitude flow in opposite directions. Thus the electric field
strength is defined as
\begin{equation}\label{eq:voltage}
E_{in}=V/b=ZJ/b,
\end{equation}
where $V$ is the line voltage, $J$ is the magnitude of current
induced in the DR element, and $Z$ is the impedance of line. The
impedance is determined at the resonant frequency
$\ae_0=d/\lambda_0$ as,
$$Z=\frac{60l\ae_0}{dC_0},$$ where $l=\pi(a_1+a_2)/2$, and
$$C_0=\frac{1}{4}\ln\left[\frac{p}{2w}+\sqrt{\left(\frac{p}{2w}\right)^2-1}\right]$$
is the capacity in free space per unit length of line, $p=a_1-a_2$.
In view of the small size of the translation cell of the array in
comparison with the wavelength, we suppose that the nonlinear
substrate remains to be a homogeneous dielectric slab under
intensive light. Thus, from this model it is possible to evaluate
the relation between the inner electric field strength and the
current magnitude $\bar{J}$ averaged along one half-arc of the
DR-element.

Next, in the \textit{nonlinear regime}, we suppose that the
structure's substrate is a material whose absorption property depends
on the intensity of the electric field
$I_{in}=(1/2)n\varepsilon_0c|E_{in}|^2$ inside it. Here the complex
refractive index $n=\sqrt{\varepsilon}$ of the substrate is defined
in the form \cite{boyd}
\begin{equation}\label{eq:susceptibility}
n=n'+in''=n'+\frac{i}{2}\frac{c}{\omega}\alpha,
\end{equation}
where $\alpha$ is an absorption index. Thus the permittivity
$\varepsilon$ of the substrate can be found as a function of
$\alpha$, $\varepsilon=\varepsilon(\alpha)$.

As a saturable absorbing material for the structure's substrate we
consider chalcogenide glasses doped with quantum dots of other
elements such as Ge, As and Sb due to their wide availability for
the mid-IR range \cite{malyarevich, zakery}. It is known that the
systems based on quantum dots have discrete energy level. To
describe the effects of saturation in such systems a two-level model
is widely used. Further, the simple two-level model is developed in
three directions. The first direction consists in use of the
three- and four-level models, which is important in studying the
process of creating a population inversion in lasers.
The second direction of the model development is taking into
account the inhomogeneous broadening of spectral lines due to the
fact that every quantum structure element (molecule, ion, quantum
dot, etc.) is in the fields of its neighbors, which may be
different. Finally the third direction is considering the
dynamic Stark shift of the energy levels of molecules under the
action of light. As usual, the two-level model is preferable when
studying bistable devices because in many cases it gives good
agreement with experiments \cite{malyarevich, zakery} and due to its
simplicity.

Thus to describe the dependence of permittivity $\varepsilon$ on the
absorption index $\alpha$  we use a two-level model of a fast
relaxing absorber whose absorption index $\alpha$ is described by
the formulae \cite{butylkin-1989}
\begin{equation}\label{eq:absorption}
\alpha(\omega,
I_{in})=\frac{0.25\alpha_0\gamma^2}{(\omega-\omega_0)^2+0.25\gamma^2(1+I_{in}/I_{sat})},
\end{equation}
where $\alpha_0$ is the weak-field absorption index, $\gamma$ is the
weak-field absorption line broadening, $I_{sat}$ is the saturation
intensity and $\omega_0$ is the frequency of the center of the
spectral absorption line. The absorption curve defined by equation
(\ref{eq:absorption}) has a Lorentzian shape, and when
$I_{in}=I_{sat}$ the absorption index is reduced to half of the peak
of the absorption line. Here the terms describing the dynamic Stark
shift are discarded under an assumption that the effect of the
dynamic Stark shift occurs under light intensities, much larger than
those required for saturation of absorption \cite{butylkin-1989}. The
absorption line broadening depends on the inner field intensity as
$\gamma\sqrt{1+I_{in}/I_{sat}}$. In the center of the spectral line,
the absorption index is $\alpha(\omega_0,
I_{in})=\alpha_0/(1+I_{in}/I_{sat})$, whereas on the distant sides
of the absorption line, $\omega-\omega_0\gg\gamma$, the value of
$\alpha(\omega, I_{in})$ is practically the same as $\alpha(\omega,
0)$, i.e., under a certain intensity level, the most pronounced
decreasing of absorption is reached in the center of the line. Hence it
follows that the best way to obtain strong interaction of the
metamaterial with a nonlinear substrate lies in tuning the
trapped-mode resonant frequency to the frequency of the center of
the spectral line of a saturable material. This fact is also proved
with experimental results of the QD's luminescence enhancement in
the plasmonic metamaterial \cite{tanaka}.

As the absorption index $\alpha$ is a function of the inner field
intensity $I_{in}$, so, too, is the permittivity of substrate,
$\varepsilon=\varepsilon(I_{in})$. According to
equation~(\ref{eq:voltage}), the inner field intensity is a function
of the averaged current magnitude, $I_{in}=I_{in}(\bar J)$. Thus the
nonlinear equation on the averaged current magnitude in the metallic
pattern can be obtained in the form \cite{tuz}
\begin{equation}\label{eq:nonlinear}
\bar J=\tilde A \cdot F_{\bar J}(\omega, \varepsilon(I_{in}(\bar
J))),
\end{equation}
where $\tilde A$ is a \textit{dimensionless} coefficient which
depicts how many times the incident field magnitude $A$ is greater
than 1~V~cm$^{-1}$. Thus, the magnitude $A$ is a parameter of equation
(\ref{eq:nonlinear}), and, at a fixed frequency $\omega$, the
solution of this equation is the average current magnitude $\bar J$
which depends on the magnitude of the incident field ($\bar J=\bar
J(A)$).

Further, on the basis of the current $\bar J(A)$ found by a
numerical solution of equation (\ref{eq:nonlinear}), the
permittivity of the nonlinear substrate
$\varepsilon=\varepsilon(I_{in}(A))$ is obtained and the reflection,
transmission and absorption coefficients are calculated as functions
of the frequency and magnitude of the incident field.

\section{Numerical results and discussion}

\begin{figure}[h]
\begin{minipage}[h]{0.5\linewidth}
\center{\includegraphics[width=1\linewidth]{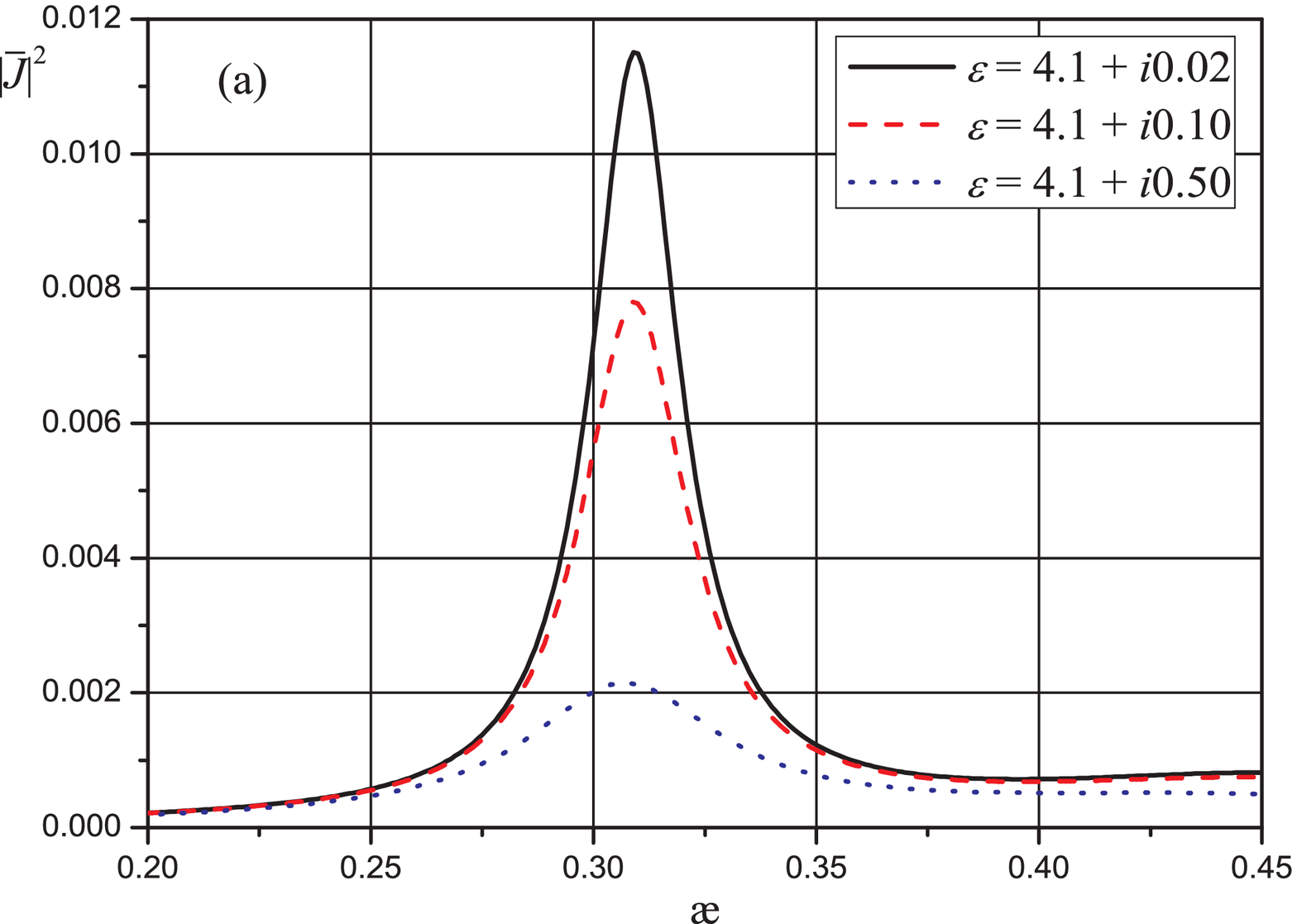}} \\
\end{minipage}
\hfill
\begin{minipage}[h]{0.5\linewidth}
\center{\includegraphics[width=1\linewidth]{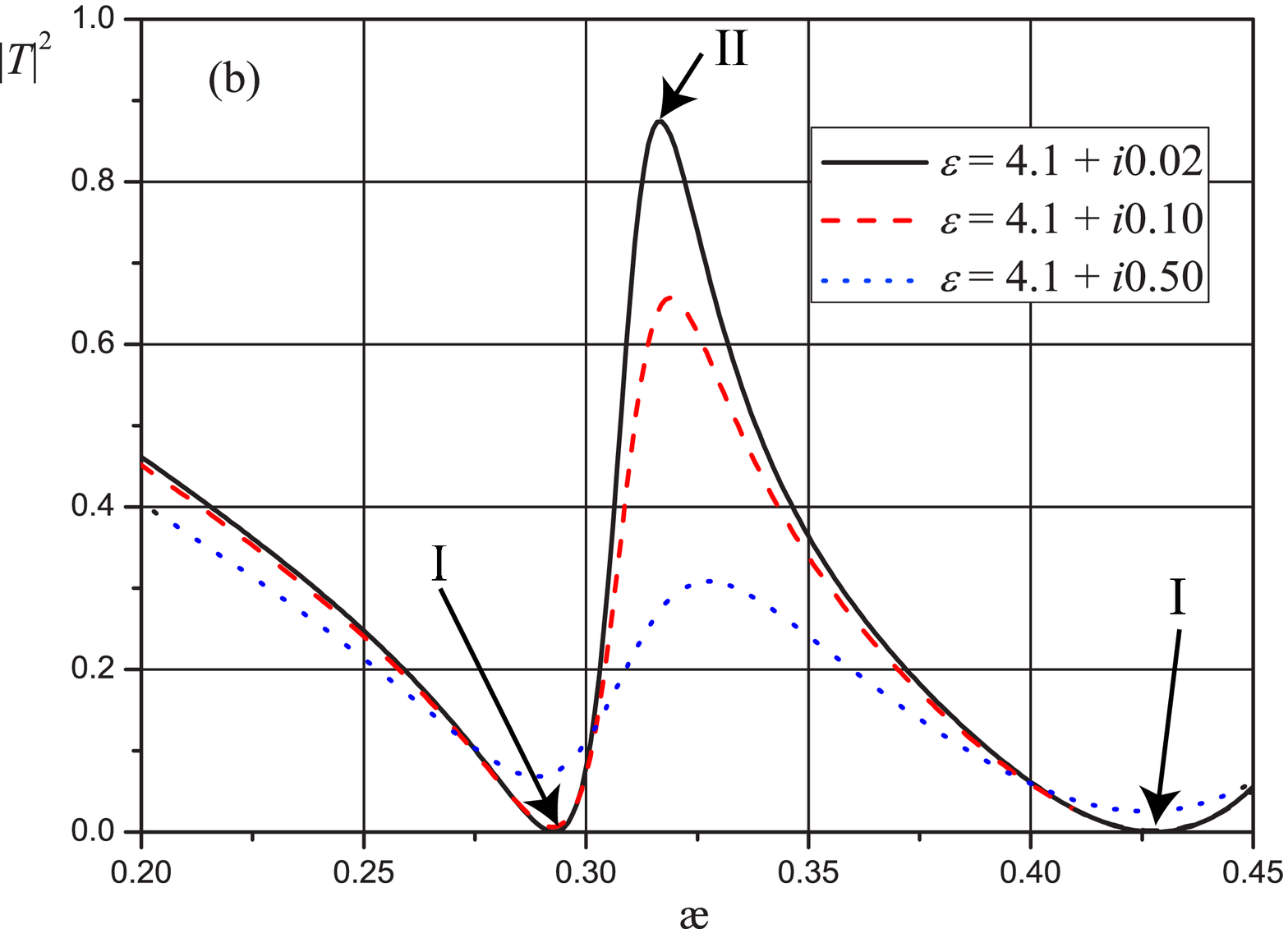}} \\
\end{minipage}
\vfill
\begin{minipage}[h]{0.5\linewidth}
\center{\includegraphics[width=1\linewidth]{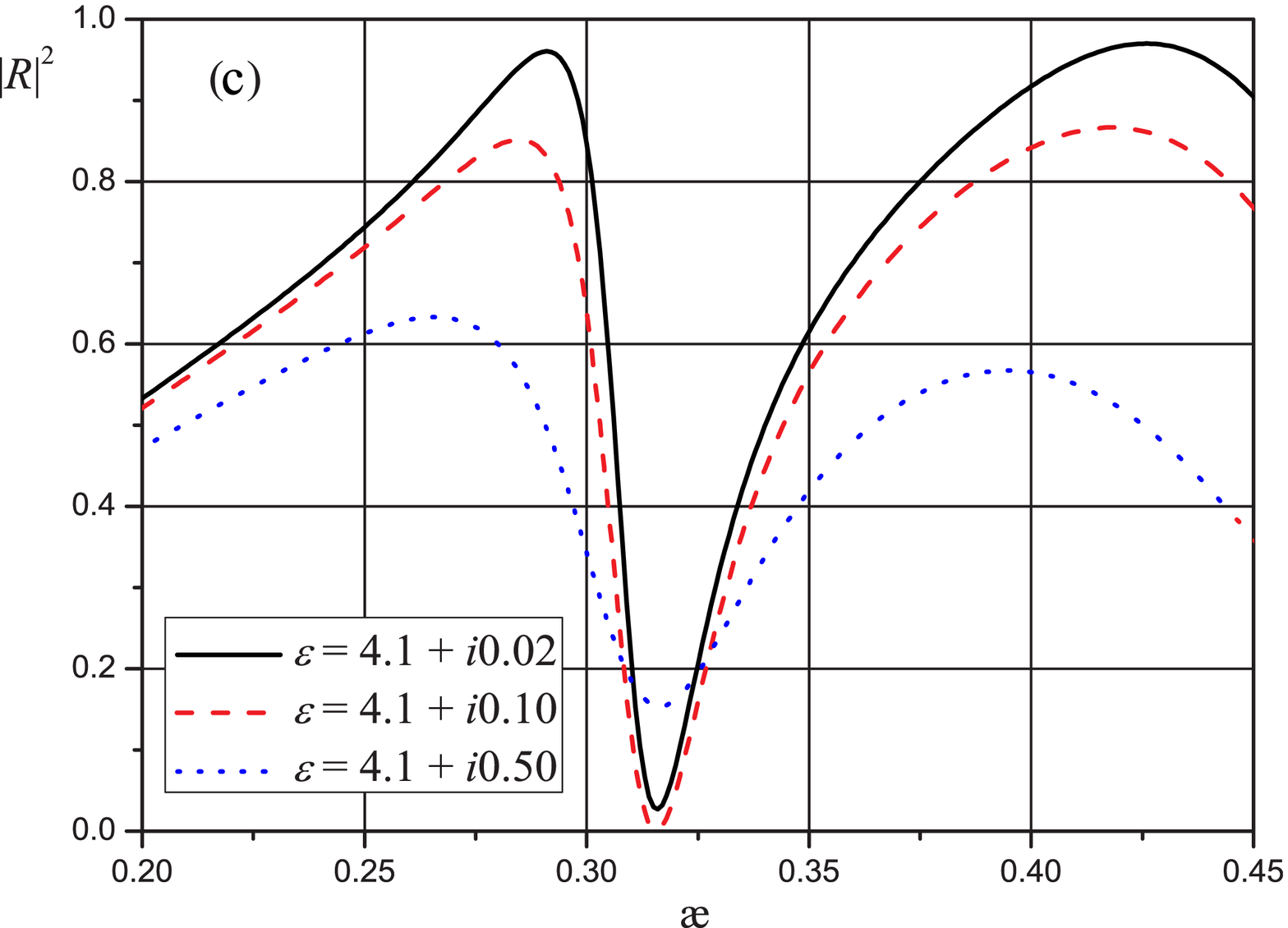}} \\
\end{minipage}
\hfill
\begin{minipage}[h]{0.5\linewidth}
\center{\includegraphics[width=1\linewidth]{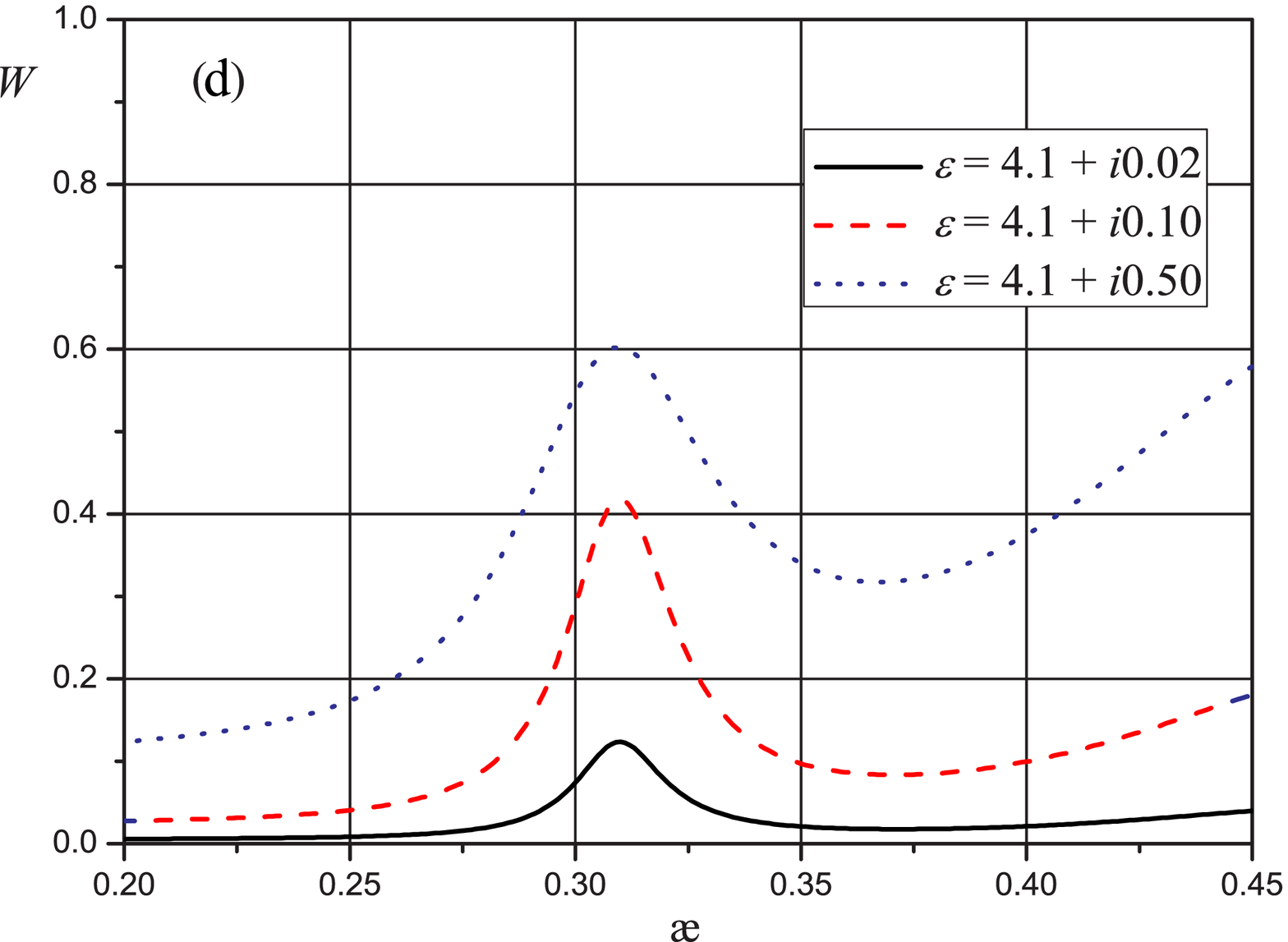}} \\
\end{minipage}
\caption{Frequency dependences ($\ae=d/\lambda$) of magnitude
squares of the average current (in a.u.) (a), transmission (b),
reflection (c) coefficients, and absorption coefficient (d) in the
case of linear regime at a low intensity of the incident field.}
\label{fig:fig2}
\end{figure}

Typical frequency dependences of the current magnitude, reflection,
transmission and absorption coefficients calculated in the linear
regime under the assumption of a nondispersive substrate with the
method of moments are given in figure.~\ref{fig:fig2}. One can see that
a sharp resonance appears in the band where the maximum of current
magnitude occurs. At this frequency band, the induced currents in
the inner and outer rings oscillate in opposite directions. An
electromagnetic field of two closed strips of the rings is similar
to the nonradiating field of a double-wire line of resonant length.
Thus, the oppositely directed but almost equal currents of the
inner and outer rings of the array yield an electromagnetically trapped
mode. The scattered field produced by such a configuration of current
is very weak in comparison with an intensity of the stored field, and,
as a consequence, the coupling of the planar metamaterial to free
space is small and, therefore, radiation losses are reduced, which
ensures a high quality factor resonant response. Evidently, as the
value of Joule losses in the metamaterial increases, the magnitude
of current and quality factor of the resonance decrease, but,
nevertheless, the resonance remains to be well observed
\cite{kawakatsu}. By this means, we suppose that, near the
frequency of trapped-mode excitation, such a metamaterial will
strongly react on the intense incident wave when the substrate is
made of a nonlinear dielectric even in the case when the dielectric
is dissipative. Also, in figure~\ref{fig:fig2}(b)
(figure~\ref{fig:fig2}(c)), one can see the evolution of approximately
\textit{symmetric} close to \textit{Lorentzian} profile of the
transmission (reflection) resonance to an \textit{asymmetric Fano}
one \cite{joe,miroshnichenko,lukyanchuk} when the losses decrease
which is very suitable to realize abrupt switching in the nonlinear
regime.

\begin{figure}
\centering
\includegraphics[width=0.6\linewidth]{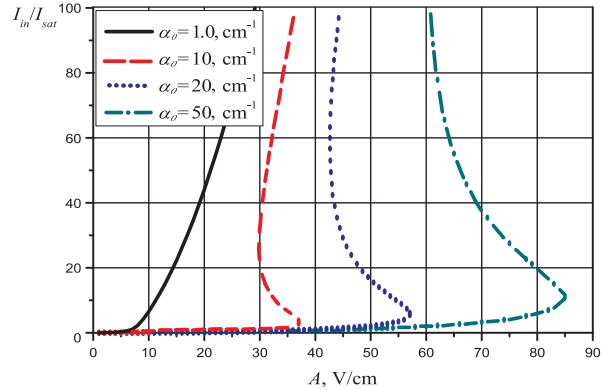}
\caption{(Color online)  Dependences of the inner intensity versus
the incident field magnitude in the case of the nonlinear
permittivity of the substrate. For this and further calculations, we
use the experimental results obtained in \cite{malyarevich}. Thus,
the saturation intensity is chosen to be $I_{sat}=60$~kW/cm$^2$,
which is typical for the phosphate glasses doped with 6-nm PbS QD's
in the IR-range, $\tilde\gamma=\gamma d/(2\pi c)=1$,
$n'=\sqrt{4.1}$. The dimensionless frequency $\ae$ corresponds to
the trapped-mode resonance and to the peak of the absorption line,
$\ae=\ae_0=\omega_0d/(2\pi c)=0.3$.} \label{fig:fig3}
\end{figure}

Thus, in the case of the array placed on the substrate whose complex
permittivity depends on the field intensity, the dependences of the
inner intensity versus the incident field magnitude
$I_{in}=I_{in}(A)$ have a typical form of hysteresis loops
(figure~\ref{fig:fig3}). The origin of the absorption bistability is
well studied in the theory of Fabry-Perot resonators \cite{gibbs},
which is also applicable to the nonlinear metamaterial. In
accordance with this theory, the nature of the hysteresis curve can
be cleared from the next considerations. A slight increase in the
incident field intensity causes a slight decrease in the
metamaterial substrate absorption which, in turn, permits the
pattern to accept more power. This added power further saturates the
saturable absorber which again permits more power to be coupled in
the pattern. This process is cumulative, with the result that the
change in the power level in the system exceeds the original small
change in the incident intensity. Thus, there is a threshold level
at which the system is unstable and abruptly switches states. A
similar process is involved as the input intensity is reduced,
causing the system to switch abruptly from a low attenuation state
to a high attenuation state at some lower threshold level.

From \cite{szoke} the condition of achieving completely absorption
bistability is also known, $\alpha_0L/T>8$, where $L$ and $T$ are
the length and transmission coefficient of the Fabry-Perot
resonator, respectively. From this condition it follows that the
switching threshold essentially depends on the value of the
absorption index $\alpha_0$. Thus, from our calculation
(figure~\ref{fig:fig3}) it is clear that, for the chosen metamaterial
parameters, the switching appears only when $\alpha_0$ is
sufficiently high since the substrate is thin.

Nevertheless there is an important distinctive feature of the
studied structure in contrast to the Fabry-Perot one. This
peculiarity is that the resonance in the Fabry-Perot cavity has the
Lorenzian-shape whereas the trapped-mode resonance in the
DR-structure has the Fano-shape, and the latter one is characterized
by an asymmetrical peak-and-trough profile.

From the classical point of view \cite{joe}, the resonant conditions
of the studied structure can be described using two weakly coupled
harmonic oscillators, where one of them is driven by a periodic
force (see, for an example, figure~2(a) in \cite{miroshnichenko}). In
such a system, in general, the spectrum of the forced oscillator
consists of two resonances, and while the first resonance is
characterized by a symmetric profile, described by a Lorenzian
function, the second one is characterized by an asymmetrical Fano
profile. At a certain frequency, the amplitude of the forced
oscillator becomes zero, as a result of destructive interference of
oscillations from the driving force and coupled oscillator.

\begin{figure}[h]
\begin{minipage}[h]{0.5\linewidth}
\center{\includegraphics[width=1\linewidth]{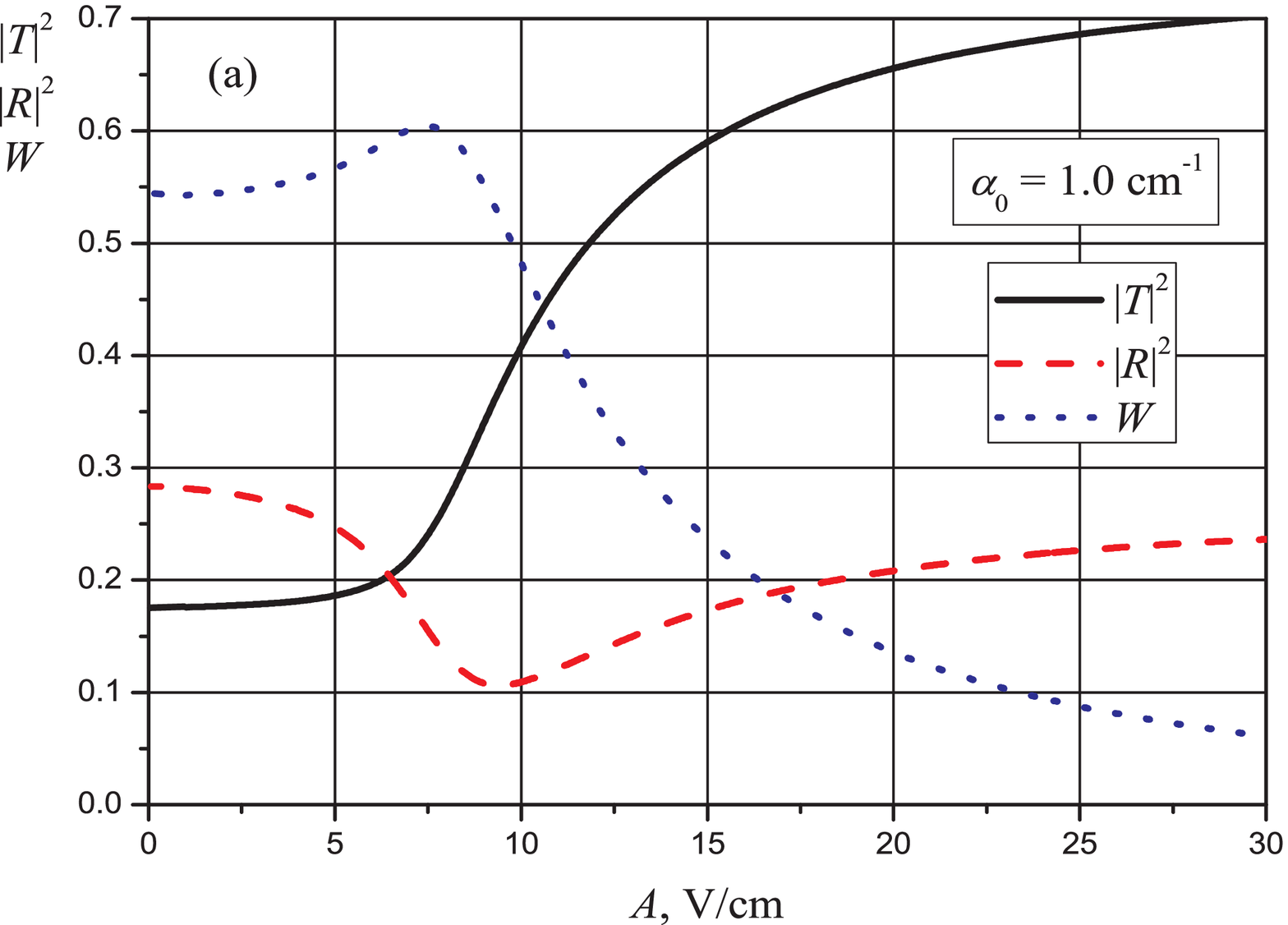}}
\end{minipage}
\hfill
\begin{minipage}[h]{0.5\linewidth}
\center{\includegraphics[width=1\linewidth]{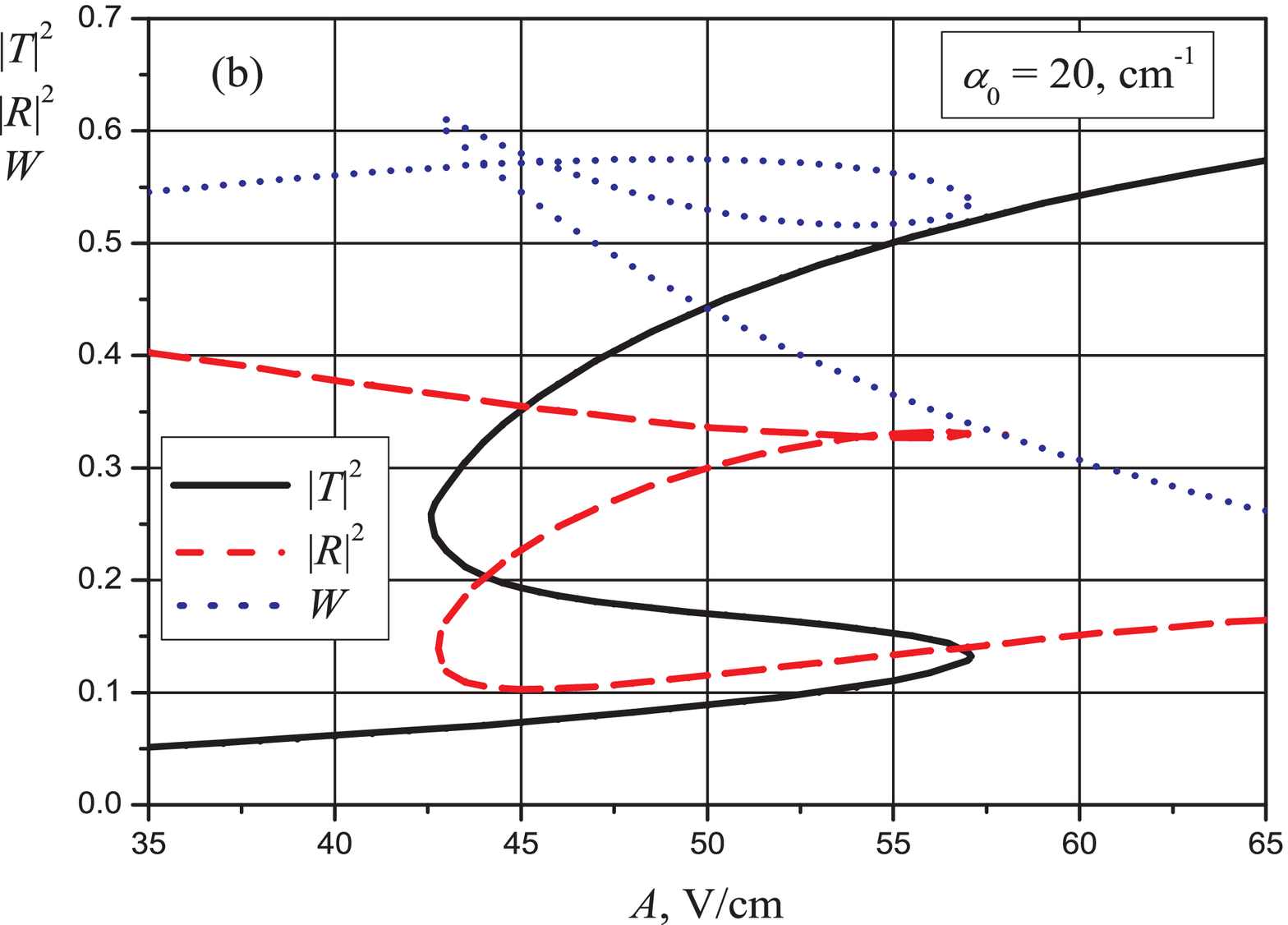}}
\end{minipage}
\caption{Dependences of the magnitudes of the reflection,
transmission and absorption coefficients versus the incident field
amplitude for different absorption index $\alpha_0$ in the case of
the nonlinear permittivity of the substrate: $I_{sat}=60$~kW/cm$^2$,
$\tilde\gamma=\gamma d/(2\pi c)=1$, $n'=\sqrt{4.1}$,
$\ae=\ae_0=0.3$.} \label{fig:fig4}
\end{figure}

In our case, the different current distributions in the metallic
pattern lead to the appearance of similar resonant conditions
(figure~\ref{fig:fig2}(b)). Thus, the resonant feature I is a result of
resonant excitation of a predominantly dipole mode (i.e., current
oscillations symmetric with respect to the symmetry axis of DR) in
either outer or inner ring and this resonance is described by a
Lorenzian function. At the feature II, both rings are excited
equally, while the induced currents in the inner and outer rings
oscillate in opposite phases, yielding an electromagnetically trapped
mode \cite{papasimakis,kawakatsu} whose resonance has an
asymmetrical profile.

Such asymmetrical spectral profile of the transmission coefficient
which vary from low to high over a very narrow frequency range
can be very useful in all-optical switchings since there are gently
sloping bands of the high reflection and transmission before and
after the resonant frequency \cite{miroshnichenko,lukyanchuk}. This
peculiarity of  spectra manifests itself in the stepwise variation
of magnitudes of reflection, transmission and absorption
coefficients calculated versus the incident field amplitude
(figure~\ref{fig:fig4}). These figures also illustrate another effect
of the shape asymmetry of the trapped-mode resonance, namely the
formation of the hysteresis with closed loops in the curves of both
reflection and absorption coefficients (figure~\ref{fig:fig4}b). In
particular, closed loop appears when the level of reflection
(absorption) before the threshold amplitude is greater then that one
after it. Thus when the hysteresis is formed, the part of the curve that
corresponds to the rising edge remains above the part of the curve that
corresponds to the falling edge.

\section{Conclusions}

In conclusion, a planar nonlinear DR-metamaterial, which bears a
high quality factor Fano-shape trapped-mode resonance is a promising
object for a realization of a polarization insensitive absorption
bistable switching. The main advantage of the studied structure is the
possibility to achieve the bistable transmission at low input
intensity, due to a large quality factor of the trapped-mode
resonance.

\ack

This work was supported by the Ukrainian State Foundation for Basic
Research, Project F40.2/037 and the Russian Foundation for Basic
Research, Project 11-02-90403.


\section*{References}


\bibliographystyle{vancouver}

\bibliography{AbsBistability}


\end{document}